# Eberhardt's inequality and recent loophole-free experiments.

Alejandro A. Hnilo

*CEILAP, Centro de Investigaciones en Láseres y Aplicaciones, UNIDEF (MINDEF-CONICET); CITEDEF, J.B. de La Salle 4397, (1603) Villa Martelli, Argentina.*
*email: ahnilo@citedef.gob.ar*
August 8th, 2016.

Recent experiments using innovative optical detectors and techniques have strongly increased the capacity of testing the violation of the Bell's inequalities in the Nature. Most of them have used the Eberhardt's inequality (EI) to close the "detection" loophole. Closing the "locality" loophole has been attempted by space-like separated detections and fast and random changes in the setting of the bases of observation. Also, pulsed pumping and time stamped data to close the "time-coincidence" loophole, and sophisticated statistical methods to close the "memory" loophole, have been used. In this paper, the meaning of the EI is reviewed. A simple hidden-variables theory based on a relaxation of the condition of "measurement independence", which was devised long ago for the Clauser-Horne-Shimony and Holt inequality, is adapted to the EI case. It is used here to evaluate the significance of the results of the new experiments, which are briefly described. A Table summarizes the main results.

**1. Introduction.**

In 1965, John S. Bell showed that the predictions of Quantum Mechanics (QM) are in contradiction with at least one of two intuitive notions, roughly speaking: *i)* the result of a measurement cannot be affected by what happens outside its past lightcone, and *ii)* the existence of a world whose properties are independent of being observed [1]. The set of these two notions is usually named Local Realism (LR). The resolution of the QM vs LR controversy is crucial to the foundations of Physics. He also proposed an experiment to decide whether QM or LR is valid in the Nature, by measuring the violation of inequalities between the statistical averages of observations performed on a spatially spread entangled state of two particles.

Due to practical limitations, the Bell's original proposal is difficult to perform. These limitations allow the existence of alternative descriptions, generally named *hidden variable theories*, which apparently violate the inequalities without violating LR. The types of practical limitations are known as *loopholes* [2]. The one often named *locality* or *freedom-of-choice* loophole arises from the possibility that information is interchanged between the remote stations where the state is observed, and the source of entangled pairs. It implies that the probability of joint detection is the product of the probabilities of detection at each station (locality) and that the hidden variables and the analyzers' settings in each station are statistically independent (*measurement independence*). To close this loophole, an unpredictable and space-like decided variation of the analyzers' settings must be achieved. The possibility that the state of the system varies in time, depending on earlier outcomes of the observations, is known as the *memory* loophole, and can be refuted through a special statistical analysis of the experimental data. The *time-coincidence* loophole is the possibility that the particle detections are shifted in time, in or out of the coincidence window. It can be disproved also by a statistical analysis, but the simplest solution is to get a pulsed source of entangled pairs and a time stamped record of the detections. The *detection* or *fair-sampling* loophole is the possibility that the particles are detected depending on the agreement of the hidden variables they carry with the analyzers' setting they find. To close it, detection efficiency higher than some threshold $\eta_{thr}$ must be achieved. For the Clauser-Horne-Shimony and Holt (CHSH) inequality, $\eta_{thr} = 2(\sqrt{2}-1) \approx 0.83$[1], for the Eberhardt inequality (EI) $\eta_{thr} = 2/3$. That's why the EI has been chosen in a series of recent optical experiments [3-6] aimed to reach the "loophole free" condition.

In the next Section, the derivation of the EI is reviewed. The QM predictions for the EI, which are not often available, are displayed. In the Section 3 a hidden variables theory, which was devised long ago to evaluate experiments using the CHSH, is adapted to the EI case. It defines an upper bound to the predictability of the analyzers' settings in order to close the locality loophole. The new experiments using the EI and their main results are briefly described in the Section 4. Another recent loophole-free experiment [7], which follows a different approach and uses CHSH, is also described. The Section 5 is the discussion of the consequences of the results of the five experiments, which are summarized in a Table at the end.

**2. The Eberhardt inequality.**

The original form of the EI is [10]:

$$N^{++}(a,b) - N^{+0}(a,b') - N^{0+}(a',b) - N^{++}(a',b') \leq 0 \quad (1)$$

where $N^{++}(i,j)$ is the number of coincidences recorded in an Einstein-Podolsky-Rosen-Bohm (EPRB) setup [1] when the analyzer's orientations settings are $\{i,j\}$, and $N^{+0}(i,j)$ ($N^{0+}(i,j)$) are the number of detections in station A (B) that do *not* produce coincidences. As the number of single detections in (f.ex.) station A is $S(a,j) = N^{+0}(a,j) + N^{++}(a,j)$, it is possible to rewrite eq.(1) as:

$$N^{++}(a,b) + N^{++}(a,b') + N^{++}(a',b) - N^{++}(a',b') - S(a) - S(b) \leq 0 \quad (2)$$

which is the Clauser-Horne (CH) inequality [8]. This is the reason why the EI is often named CH-Eberhardt inequality.

---

[1] $\eta_{thr} = 0.83$ is the value stated in the earliest description of this loophole [8]; a different approach leads to $\eta_{thr} = 1/\sqrt{2} \approx 0.71$ and even to $[2(\sqrt{2}-1)]^2 \approx 0.68$ [9]. The safest criterion is to use the most stringent condition.

Ideally, the bound is violated by some entangled states, implying that QM is not compatible with LR. In order to violate the bound in a real experiment, the detection efficiencies must be taken into account. Assuming, for simplicity, that the efficiency $\eta$ is the same for all the detectors and settings, and dividing by the total number of pairs for each setting (assumed equal) in order to get probabilities, the eq.(2) becomes:

$$\eta^2 \times [P^{++}(a,b) + P^{++}(a,b') + P^{++}(a',b) - P^{++}(a',b')] - \eta \times [P^+(a) + P^+(b)] \equiv J \leq 0 \quad (3)$$

The probability of singles is always larger than the probability of coincidences for the same setting and, to make things even worse, singles are multiplied by $\eta$ instead of $\eta^2$, so that violating the inequality seems impossible for practical values of $\eta$.

Eberhardt's brilliant idea was the use of a non-maximally entangled state. It is certainly anti-intuitive that, to test QM vs LR, a partially entangled state can be *better* than a maximally entangled one. He defined the state ($r < 1$):

$$|\psi_E\rangle = (1+r^2)^{-\frac{1}{2}}\{|x_A, y_B\rangle + r|y_A, x_B\rangle\} \quad (4)$$

whose Concurrence is $2|r|/(1+r^2)$. The probability of coincidences for this state is:

$$P^{++}(a,b) = (1+r^2)^{-1}[\cos(a)\sin(b) + r\sin(a)\cos(b)]^2 \quad (5)$$

and the probabilities of single detections:

$$P^+(a) = (1+r^2)^{-1}[\cos^2(a) + r^2 \sin^2(a)] \quad (6)$$

$$P^+(b) = (1+r^2)^{-1}[r^2 \cos^2(b) + \sin^2(b)] \quad (7)$$

Choosing the angle settings so that $\cos(a) \approx 0$ and $\sin(b) \approx 0$, the single probabilities are $\approx r^2$. From eq.(5) also $P^{++}(a,b)$, $P^{++}(a,b')$ and $P^{++}(a',b) \approx r^2$. The settings $\{a',b'\}$ are free to make $P^{++}(a',b') \approx 0$. Then:

$$J \approx 3\eta r^2 - 2r^2 \leq 0 \quad (8)$$

so that $\eta \geq 2/3$ to violate the bound, which is the well-known result $\eta_{thr} = 2/3$. Fine tuning of r and the angle settings maximizes J. Replacing the eqs.(5)-(7) into eq.(3) with $\eta=1$ gives the ideal QM prediction $J_{QM}$. It is easy to calculate, but cumbersome to display as an explicit function. An approximate expression is:

$$J_{QM} \approx \eta r^2 \quad (9)$$

F.ex, for the data in [4], the exact value ($\eta=1$) is $J_{QM} = 0.067$, not too far from $r^2 = 0.084$.

Some setups use the state $|\phi_E\rangle = (1+r^2)^{-\frac{1}{2}}\{|x_A,x_B\rangle + r|y_A,y_B\rangle\}$. The probability of a coincidence is then:

$$P^{++}(a,b) = (1+r^2)^{-1}[\cos(a)\cos(b) + r\sin(a)\sin(b)]^2 \quad (10)$$

The probability of single detections is given by eq.(6) for both stations, due to the symmetry of $|\phi_E\rangle$.

## 3. HV+DZ for the Eberhardt inequality.

In 1991, a simple hidden variables model called HV+DZ was proposed as a test bench of the significance of the results of EPRB experiments using the CH or CHSH inequalities [11]. It is based on a yes-no probability distribution of passage on the angle setting of the analyzers, see the Figure 1.

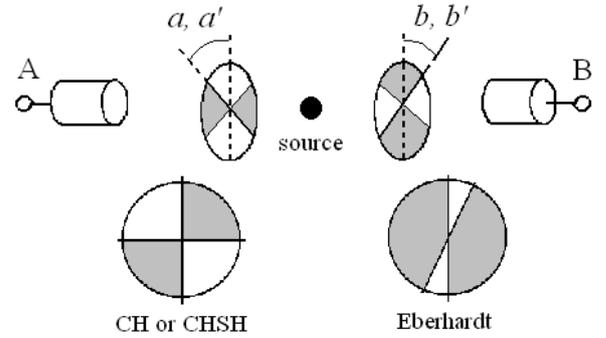

Figure 1: Each analyzer has "transparent" (white) and "reflective" (grey) regions in the space of the hidden variables. The probability of coincident detection is given by the overlap of the transparent regions. Their size is different if r=1 (CH or CHSH, $P^+= \frac{1}{2}$) or r<1 (EI, $P^+< \frac{1}{2}$).

This basic scheme saturates the inequality. It may seem that a small displacement of the transparent regions may suffice to violate it. But it is not, because the advantage obtained for some analyzers' settings is compensated by the disadvantage for others. An auxiliary set of hidden variables is then added, that define "target" detectors for the particles and corresponding displacements of the transparent areas. A critical parameter is then the probability the particles have to reach their target detectors, named $\frac{1}{2}\sqrt{a}$ ($a>1$). This correlation between the auxiliary hidden variables and the analyzers' settings means a violation of measurement independence. The experiments attempt to enforce it by random and space-like choosing of the settings. Yet, some (small) correlation may still exist due to the statistical predictability of the future settings (because of some unbalance or bias in their distribution) and the imperfections of the fast modulators and/or analyzers at each station. The problem addressed by HV+DZ is to calculate the minimum degree of correlation (hence, of deviation from strict measurement independence) necessary to reproduce the QM predictions.

The HV+DZ model was devised to study the experiment by Aspect *et al.* [12]. Later, it was used to evaluate the improvements reached by the experiment by Weihs *et al.* [13,14]. The value of the correlation ½√$a$ necessary to reproduce the ideal QM predictions for the CH and CHSH inequalities was found to be $(1/\sqrt{2} - 7/16)^{1/2} + \frac{1}{4} \approx 0.769$, or ≈0.27 above the minimum value of ½. The consequences of a deviation from perfect measurement independence were considered by several authors from different points of view [15-17]. A general approach almost halved the excess value calculated in HV+DZ, to $(\sqrt{2}-1)/3 \approx 0.14$ [15].

In few words: a relatively small relaxation of measurement independence allows reproducing the ideal QM predictions for the CH and CHSH inequalities. It is then pertinent to study how such relaxation affects the EI.

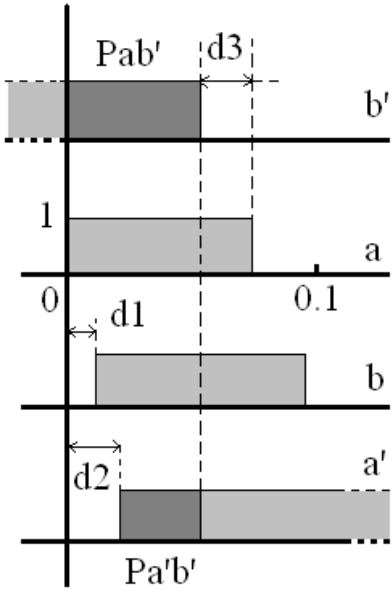

Figure 2: Scheme of the probabilities of detection for each setting and station, for the HV+DZ model adapted to the EI. The horizontal scale is approximate.

The HV+DZ model can be adapted to the EI as follows. The probabilities of detection in each station and setting, as a function of the angular variable, are given by the scheme in the Figure 2 (compare with Fig.3 in [11]). All the grey areas have height equal to 1, so that their lengths are equal to the probabilities of single passage. For example, using the data in [5] the lengths of the areas in the axes named "a" and "b" are $P^+(a) = 0.0825$ and $P^+(b) = 0.0884$. The values of $P^+(a')$ and $P^+(b')$ are irrelevant for the EI; here they are both ≈ ¼ and extend out of the figure. The overlap of two areas is the probability of the corresponding coincidence. In the Fig.2, the dark grey areas indicate $P^{++}(a,b')$ and $P^{++}(a',b')$ as an illustration. In the original HV+DZ model, the length of all the grey areas was ½, to fit the probability of single passage for a maximally entangled (r= 1) state. Note the definition of the displacements $d_j$:

$$P^{++}(a,b) = P^+(a) - d_1$$

$$P^{++}(a',b) = P^+(b) + d_1 - d_2$$
$$P^{++}(a,b') = P^+(a) - d_3 \qquad (11)$$
$$P^{++}(a',b') = P^+(a) - d_2 - d_3$$

The EI is saturated (J=0) regardless the values of the $d_j$.

Let define now an auxiliary hidden variable μ. The pairs with μ=1 have the setting {a,b} as their "target", the ones with μ=2 the {a,b'}, μ=3 the {a',b} and μ=4 the {a',b'}. Roughly speaking, each pair of emitted particles tries to reach its target by guessing the future settings when leaving the source, and also by exploiting the instrumental imperfections when the guess fails; the probability $q$ (= ½√$a$ in the notation of [11]) of reaching the target is (at first order) the sum of the probabilities of success of both strategies. Strictly speaking, $q$ is a measure of the correlation between the hidden variable μ and the settings regardless of the physical cause; $\varepsilon = q - \frac{1}{2} > 0$ is then the amount of the deviation from perfect measurement independence. The set of the $d_j$ is therefore enlarged into a set of $d_{j\mu}$ whose values are chosen to maximize the value of J:

$$P^{\mu=1}(a,b) = P^+(a) \;(\Leftrightarrow d_{11}=0)$$
$$P^{\mu=2}(a,b') = P^+(a) \;(\Leftrightarrow d_{32}=0) \qquad (12)$$
$$P^{\mu=3}(a',b) = P^+(b) \;(\Leftrightarrow d_{13}=d_{23})$$
$$P^{\mu=4}(a',b') = 0 \;(\Leftrightarrow d_{24}+d_{34}=P^+(a))$$

F.ex., for μ=1 one wants to get $P^{\mu=1}(a,b) = P^+(a)$, which is the maximum possible value for $P^{++}(a,b)$. Therefore, one must define $d_{11}= 0$, and then $P^{\mu=1}(a,b')= P^+(a) - d_{31}$, $P^{\mu=1}(a',b)= P^+(b) - d_{21}$ and $P^{\mu=1}(a',b')= P^+(a) - d_{21} - d_{31}$. In the same way, expressions for all the $P^\mu(i,j)$ are found. Note that the values of some $d_{j\mu}$ remain free. Assuming for simplicity that $q$ is the same for all μ, the observable coincidence probabilities are:

$$P^{++}(a,b) = q^2 P^{\mu=1}(a,b) + q(1-q)[P^{\mu=2}(a,b) + P^{\mu=3}(a,b)] + (1-q)^2 P^{\mu=4}(a,b)$$
$$P^{++}(a,b') = q^2 P^{\mu=2}(a,b') + q(1-q)[P^{\mu=1}(a,b') + P^{\mu=4}(a,b')] + (1-q)^2 P^{\mu=3}(a,b')$$
$$P^{++}(a',b) = q^2 P^{\mu=3}(a',b) + q(1-q)[P^{\mu=1}(a',b) + P^{\mu=4}(a',b)] + (1-q)^2 P^{\mu=2}(a',b)$$
$$P^{++}(a',b') = q^2 P^{\mu=4}(a',b') + q(1-q)[P^{\mu=2}(a',b') + P^{\mu=3}(a',b')] + (1-q)^2 P^{\mu=1}(a',b') \qquad (13)$$

Replacing the $P^\mu(i,j)$ of eqs.(12) into eqs.(13) and then into eq.(3) with η=1, the value of J according to the HV+DZ model is:

$$J_{DZ} = q^2 P^+(a) - (1-q)\{qP^+(a) + (1-2q)[d_{14} + d_{22} + d_{33} - (d_{12} + d_{21} + d_{31})]\} =$$
$$= q^2 P^+(a) - q(1-q)P^+(a) - (1-q)(1-2q)\{P^+(a) + P^+(b) - P^{\mu=4}(a,b) - P^{\mu=2}(a',b) - P^{\mu=3}(a,b') + P^{\mu=1}(a',b')\} \qquad (14)$$

note that $q = \frac{1}{2} \Rightarrow J_{DZ} = 0$, as it must be. The factor between keys in the second equality has the form of an EI but with the opposite sign and coincidence probabilities for *non* target settings. It is convenient to define this factor as (-J'), so that eq.(14) is written:

$$J_{DZ} = q^2 P^+(a) - q(1-q) P^+(a) + (1-q)(2q-1)(-J')  \quad (15)$$

The numerical value of J' depends on the precise choosing of the $d_{j\mu}$ that remained free. It is always J'≤0 because this choosing defines a LR hidden variables theory, which necessarily holds to the EI. The last term in eq.(15) is hence positive, or zero. Then:

$$J_{DZ} \geq P^+(a) \times (\varepsilon + 2\varepsilon^2) \quad (16)$$

so that the EI is violated by *any* value of *q* larger than ½. The same is valid for the original HV+DZ case (r=1) and CHSH:

$$S = 2 \times (1 + \varepsilon + 2\varepsilon^2) \quad (17)$$

A different situation is faced if J is required not to merely violate the bound, but to *fit* the QM prediction $J_{QM}$. For a choosing of the remaining $d_{j\mu}$ such that -J' = $2P^+(a) + P^+(b) \approx 3P^+(a)$ then, from eq.(15):

$$q_{QM} \approx 1 - \tfrac{1}{2} [1 - J_{QM} / P^+(a)]^{\frac{1}{2}} \quad (18)$$

For the ideal values in [5], $q_{QM} \approx 0.78$. Compare with $q_{QM} = 0.769$ required by the original HV+DZ model to fit S= 2√2. Replacing measured values of J and $P^+(a)$ in eq.(18), the minimum correlation $q_m$ the HV+DZ needs to reproduce the observations is obtained.

In summary: the counterexample provided by the HV+DZ model shows that the EI bound is violated as soon as a relaxation of measurement independence is allowed, even if η=1. The same holds for the CHSH. The ideal values $J_{QM}$ and S=2√2 are reached by $q_{QM}\approx$ ¾ in both cases. Hence, the capacity to discriminate QM from LR is, in this sense, the same by the EI and the CHSH. The only difference is the smaller value of $\eta_{thr}$ of the EI. Yet, recall that it rapidly increases in the presence of background noise.

## 4. Brief description of the new experimental results.

*4.a Definition of relevant parameters in common.*
Three experiments (4.b, 4.c and 4.d) are based on the use of Transition Edge Sensors (TES) [18] operating at cryogenic temperatures. These photon detectors have a measured efficiency up to 98%, and have been crucial to close the detection loophole.

In order to close the memory loophole, a refined statistical method is performed, which was firstly developed in [19]. The idea is to calculate the maximal probability (what is called the *p-value*) that the observed outcome has been produced by a statistical fluctuation under the conditions to be disproved, in this case, LR. This type of test was formalized for the first time in [20,21]. Other methods to calculate the p-value were developed in [22,23], to include a bias in the setting distributions (which is a common practical imperfection) for both the CHSH and EI. The predictability of the settings is taken into account in the recently published paper [24], allowing for two different cases: perfect predictability in some trials, or else, constant average predictability during the whole run. The latter can be considered as a different approach to the situation addressed by HV+DZ. As it will be seen, the two approaches lead to the same conclusion. The p-value is used in the experiments 4.d to 4.f to close the memory loophole, and in 4.d and 4.e also as an alternative to the calculation of the standard error of J to quantify the reliability of the results. A main practical concern is finding the value of the cut-point that provides a statistically significant p-value. This can be done without assuming any *a priori* distribution, following a special procedure that is detailed, f.ex., in the supplemental material of experiment 4.e. Note that the value $q_m$ obtained from HV+DZ is not the consequence of a statistical fluctuation, but of a systematical behavior at the hidden variables level. It is valid even in the limit of an infinite statistical set. In general, the p-value and $q_m$ define different and complementary bounds.

In principle, the correlation between the hidden variable μ and the analyzers' settings can have two causes: the predictability of the settings and the instrumental imperfections of the settings' realizations. In the initially studied experiment of Aspect *et al.*, the latter was mainly determined by the contrast of the acousto-optical modulators that deflected the beams towards the different fixed settings. In the recent experiments, it is related with the latency time of the electro-optical modulators (EOM) and the errors in the time stamping devices. For simplicity, I take into account here the first cause only. The predictability of the settings is given by the random number generators (RNG). Using physical models of the random processes and measurements of the RNG outputs, estimates of the predictability in each setup, $q_{set}$, can be developed. In what follows, the values of $q_{set}$ are the ones estimated by the authors of the experiments. To close the locality loophole, the value of $q_{set}$ must be smaller than the value $q_m$ the HV+DZ needs to reproduce the observed value $J_m$, or $q_{set} < q_m$.

Other relevant parameters are the independently measured efficiencies of the detectors $\eta_{meas}$. Inserted in eq.(3), they provide a "first-order" corrected value $J_{corr}$. One expects $J_{corr}/J_m \approx 1$. Conversely, $\eta_{eq}$ is the efficiency value that, inserted into eq.(3) with the ideal values of the probabilities, reproduces $J_m$. It is a compact way to take into account the experimental imperfections. Ideally, $\eta_{eq} \approx \eta_{meas}$. The criterion to close the detection loophole is $\eta_{eq} > \eta_{thr}$.

Be aware that the CHSH parameter S is used instead of J in the experiment 4.f. Also, note that the coincidence rate depends on the angle settings, which are different in each experiment, so that a direct comparison is impossible. Hence, the displayed values just give a rough idea of the rate of detected particles and of the contrast of the coincidence vs. angle curves in each case. The parameters' values are summarized in the Table at the end.

*4.b Giustina et al., 2013.*

A CW laser diode at 405 nm pumps a ppKTP-II crystal placed inside a Sagnac interferometer. The insertion of additional crystals on the pump beam prepares the state $|\psi_E\rangle$ with r≈0.3. The pairs of photons at 810 nm pass analyzers with fixed settings, are selected with interferential filters, focused into single mode optical fibers and detected with TES. The time of detection of each photon is stored in a time-tagged file. Each setting is left fixed during 300 s.

The measured value of the violation of the EI is $J_m = 5.24 \times 10^{-3} \pm 8 \times 10^{-5}$, while the ideal value is $J_{QM} = 7.01 \times 10^{-2}$. The difference is explained by taking into account several practical imperfections [25]. Without going into the details, note that $J_{corr} = 8.53 \times 10^{-3}$, which is reasonably close to $J_m$. Conversely, the equivalent efficiency is $\eta_{eq} = 0.745$. Hence, assuming a small deterioration of the average efficiency suffices to explain the difference between $J_m$ and $J_{corr}$. This feature is found even in the other experiments using EI, where $J_{corr}/J_m$ is larger (see the Table). The reason why the value of the efficiency is so critical to the value of J is a consequence of the EI, which deals with numbers that are all close to zero.

The measured value of $P^+(a)$ is 0,063 reasonably close to the ideal 0,088. From eq.(18), $q_m \approx ½ + 2 \times 10^{-2}$. In this setup and the next the settings are fixed, so that $q_{set} = 1 > q_m$. But, recall that these experiments were *not* aimed to close the locality loophole.

*4.c Christensen et al., 2013.*
The third harmonic of a mode-locked Nd:YAG laser (5 ps at 120 MHz) is used to pump a pair of crossed BBO-I crystals. Additional crystals prepare the state $|\phi_E\rangle$ with r= 0.26. The pairs of photons at 710 nm are filtered with single mode optical fibers and detected with TES cooled at 100 mK. The TES have a jitter about 1 µs, so that they are unable to resolve the time between the mode-locking pulses. The pump beam is modulated with a Pockels cell to produce bursts 1 µs long (or 240 mode-locking pulses), separated by 40 µs. The time of detection of each photon is saved in a time-tagged file. The settings are randomly changed with a periodicity of 1 s. This is not done to close the locality loophole (the period is too long for that), but to avoid any instrumental drift that may produce a spurious violation of the EI.

The measured value $J_m = 5,4 \times 10^{-5} \pm 7 \times 10^{-6}$ is three orders of magnitude smaller than the ideal $J_{QM} = 5,49 \times 10^{-2}$. Yet, $\eta_{eq} = 0.71$ suffices to fit the two values. In this experiment and in the previous one $\eta_{eq} > \eta_{thr} = 2/3$, so that the detection loophole is successfully closed in both cases. In this experiment the CHSH parameter is also measured (after adjusting r = 1) and the excellent value $S_{CHSH} = 2.827 \pm 0.017$ is obtained.

The measured value of $P^+(a)$ is $\approx 1.69 \times 10^{-3}$, much smaller than the ideal 0,067. However, as $J_m$ is also small, $q_m$ is only $\approx ½ + 8 \times 10^{-3}$.

*4.d Giustina et al., 2015.*
The source is the same than in 4.b (r = -0.29) but the 405 nm pump laser diode is now modulated to emit pulses 12 ns FWHM at 1 MHz repetition rate. An output signal from the laser synchronizes the measuring process and defines the "natural time" a valid photon is expected to arrive to the stations. The entangled photons are spectrally and spatially filtered by focusing into single mode optical fibers, which transport them to the stations of observation. The stations are separated ≈58m, with the source near the middle point. In each station, EOM are driven to change the angle settings of the analyzers. The settings are decided by identical RNG, one in each station. The RNG are based on laser phase diffusion, and produce raw series of random bits at 200 MHz speed. At the time an output is required, only the most recent raw bits are chosen to run a parity calculation to decide the measurement setting. This is to make sure that the definitions of the settings are space-like separated. The process of choosing and driving the EOM is completed in only 26 ns, shorter than the distance from each station to the source (≈87 ns). The predictability of this process is estimated smaller than $½(1 + 2.4 \times 10^{-4})$. The photons are detected with TES with independently measured efficiencies $\eta_A = 0.786$ and $\eta_B = 0.762$.

All the relevant data are saved in time-tagged files. The digitizer operates in a triggered mode starting with a photon detected during the natural time (which is defined by the signal coming from the pump laser), but it also records 1024 ns before and after the trigger at a sample rate of 250 MHz, or 4 ns time resolution. This is far more than enough, taking into account the jitter of the detectors. One limitation is that the digitizer requires 2.176 µs to re-arm after a trigger. This means that it is blind to the next two or three pump pulses. Yet, this is not a serious limitation, for the probability of producing one photon per pulse is low, to keep the number of accidental coincidences small [26].

Data are recorded during 4.8 hours, in three blocks of one, one and 2.8 hours. The experiment stops when a real-time running check of entanglement indicates that the setup is drifting out of alignment. Using the second block of data, the value $J_m = 7.27 \times 10^{-6}$ is obtained, four orders of magnitude smaller than the ideal one.

The measured value of $P^+(a)$ is not directly provided. As the singles in station "1" must be independent of the setting in station "2", I estimate $P^+(a)$ from the sum $N_{11}^{++} + N_{12}^{+0} = 141439 + 67941$ (see supplementary material in [5]) and dividing by the average of the number of trials in each case, or (875683790+875518074)/2, then $P^+(a) \approx 2,4 \times 10^{-4}$. It agrees with the value $2,2 \times 10^{-4}$ obtained using $\eta_{eq}$, but it is far from the ideal 0,083. Anyway, $q_m = ½ + 7,6 \times 10^{-3}$ is larger than $q_{set} = ½ + 1.2 \times 10^{-4}$, so that the locality loophole is closed. If the ideal value of $P^+(a)$ were used instead, $q_m = ½ + 2,2 \times 10^{-5}$ and the loophole would not be closed. The p-value defined in this experiment takes into account the predictability of the settings. It is measured smaller than $3.74 \times 10^{-31}$, hence closing both the memory and the locality loopholes.

Background photons raise the value of $\eta_{thr}$. Here, the electrical signal produced by the TES is digitized and its shape is used to discriminate detections of valid photons at 810 nm from photons of lower energy coming from blackbody radiation. However, remaining background is claimed, together with imperfect state purity, for the large difference between $J_m$ and $J_{corr}$. I have found no reported estimation of the remaining background level or of the increase of $\eta_{thr}$.

*4.e Shalm et al., 2015.*
A mode-locked Ti:Sapphire laser emits ps pulses at a repetition rate of 79.3 MHz at 775 nm. It is split in two beams with orthogonal polarizations to pump a ppKTP crystal. These two beams are inserted into a polarization Mach-Zehnder interferometer that allows preparing the state $|\phi_E\rangle$ (r =0.287). The down-converted photons are in the communications band, which is an attractive feature of this setup. They are sent to the observation stations via optical fibers. The mode-locked pulses are used as a clock to synchronize the measurements. They are detected at the laser's output with a fast photodiode, one each of 800 is picked out, and the electrical signal is sent to the stations. Once per second, a signal from the GPS helps to prevent any slow drift between the time tags in each station during an experimental run (30 min).

The source and the stations are positioned at the vertices of a nearly right-angle triangle. Each station is at about 130 m from the source, and the straight line distance between the stations is 184.9 m. While the photons are in flight, the electrical signal from the pump laser triggers a RNG, to choose a measurement setting. This occurs at a rate 79.3 MHz/800 = 99.1 KHz. At each station, the photons pass through an EOM and are detected by a superconducting nanowire single photon detector (SNSPD) with an efficiency of 91% [27]. The detected signal is saved in a time-tagger with a 10 MHz clock. The process is completed before any information from the other station may arrive at luminal speed.

However, the EOM remain on the same state for about 200 ns, or 15 mode-locked pulses. Only the pulses at the center of this set fulfill the condition that the setting is completed before the photon arrival. Then, there is a compromise: taking into account only pulse #6 to get the best condition of space-like separation and scarce statistics, or else, adding neighboring pulses to improve the statistics at the cost of relaxing the condition of space-like separation. The number of pulses taken into account into the statistics (which can be decided after the experiment has finished) is therefore a crucial parameter of this experiment. A table of p-values for different number of pulses and predictability excess of the RNG is provided. Excepting for the extreme cases, the memory loophole is clearly closed.

As in 4.d, the probability per pulse of generating a pair is very low ($\approx 5\times 10^{-4}$), so that the chance of getting two events inside the same time window is negligible (<1%). Yet, if the aggregate pulses do not fulfill the condition of being space-like (even if less than one event per time window is recorded in the average) the test becomes, of course, unreliable.

The RNG in this setup deserve a special comment. There are three in each station. Two of them are relatively common: one is based on measuring optical phase diffusion in a gain-switched laser, the other one on sampling the amplitude of an optical pulse at the single-photon level. The third one is unusual: it produces a bit from XORing the digitized version of popular movies (one different in each station) with the digits of $\pi$. I had discussed the issue of the RNG in [11], and found that there were two possible sources of unpredictable results. One was a quantum state that projected into orthogonal bases. The other one was a series of random numbers stored in a computer. There was always an untestable hypothesis involved: *i)* the projection of the quantum state was uncorrelated with the hidden variables carried by the pair, or *ii)* the source of pairs was unable to read the memory of the computer. In this experiment, both possible sources are available and their outputs are scrambled. The predictability of this (in my opinion, "ultimate") source is estimated smaller than $\frac{1}{2} + 10^{-4} = q_{set}$.

The value $J_m = 1.41\times 10^{-5}$ is obtained for 5 aggregate pulses (around the optimal pulse #6) more than three orders of magnitude smaller than the ideal value. The p-value is $5.9\times 10^{-9}$, closing the memory loophole. The p-value obtained using the method developed in [20-21] is comparable. If an excess predictability of the RNG even 15 times larger is allowed, the p-value raises to only $2.3\times 10^{-7}$. The result for 7 aggregate pulses is worse ($2\times 10^{-7}$ and $9.2\times 10^{-6}$ respectively), in spite of fulfilling the space-like condition and the improved statistics. This is probably due to instabilities of the voltage applied to the EOM near the moment a setting change is made.

The measured value of $P^+(a)$ is $2.17\times 10^{-4}$, in the order of the value estimated in 4.d and, once again, far from the ideal 0,081. Then $q_m = \frac{1}{2} + 1{,}6\times 10^{-3} > q_{set}$ so that the locality loophole is closed. If the ideal value of $P^+(a)$ were used instead, $q_m = \frac{1}{2} + 4{,}4\times 10^{-5}$ and the loophole would not be closed.

The effect of background counts coming from blackbody radiation and room light is carefully taken into account. In order to reduce their number, the only events considered are those that occur within a window of 625 ps (at station A) and 781 ps (B) around the natural time. The probability of observing a background count during the natural time is found to be $8.9\times 10^{-7}$ (A) and $3.2\times 10^{-7}$ (B). These numbers raise $\eta_{thr}$ from 2/3 up to 0.725, marginally larger than $\eta_{eq} = 0.715$. Therefore, this experiment is in the limit of closing the detection loophole.

*4.f Hensen et al., 2015.*
This experiment is very different from the previously described ones. In particular, it does not use the EI, but the CHSH.

The entangled particles to be detected in this experiment are not photons, but the electronic spins associated with single nitrogen vacancy defects (NV) in diamond chips. The spin orientation can be handled by applying RF signals. The efficiency its state can be measured is close to 100%. The time required to measure the state of a NV is relatively long, what makes necessary a distance between the stations longer than in the previous experiments, in order to ensure that all the setting choices and measurements are space-like isolated. The NV are then placed in stations A at 493m and B at 818m from the "source" station (see below), the distance between A and B is 1280m. Each spin is entangled with an emitted photon, which is inserted into an optical fiber and sent to the source station. There, the two photons are subjected to a Hong-Ou-Mandel measurement. If coincident photons are recorded at the two output ports of a beam splitter, then the biphoton state is $|\psi^-\rangle$ ($|\psi_E\rangle$ with r= -1) and then the two remote NV are also in the state $|\psi^-\rangle$. This is because of the phenomenon of *entanglement swapping*, which is based on the equality involving the states of the Bell's basis:

$$|\psi_{1A}^-\rangle|\psi_{B2}^-\rangle = \tfrac{1}{2}(|\psi_{12}^+\rangle|\psi_{AB}^+\rangle - |\psi_{12}^-\rangle|\psi_{AB}^-\rangle - |\phi_{12}^+\rangle|\phi_{AB}^+\rangle + |\phi_{12}^-\rangle|\phi_{AB}^-\rangle) \quad (19)$$

where 1,2 are the photons and A,B are the NV.

This setup is the closest to the original Bell's proposal, for the photons detected at the source not only prepare the NV state, but they also play the role of an "event-ready" signal heralding that an entangled state (of the two NV) is available for measurement.

Yet, the whole process occurs rarely. The probability of entanglement generation per attempt is estimated $6.4 \times 10^{-9}$, or slightly more than one event-ready signal per hour. Once an event-ready signal from the source is recorded, the detection probability of the NV is almost 100%, closing the detection loophole ($\eta_{thr} = 0.83$ here). The whole experiment ran 245 trials during a total measurement time of 220 h. The measured CHSH parameter is $S_m = 2.42$ (noteworthy, it is higher than the previous estimation, $S_{corr} = 2.30$). The HV+DZ model needs $q_m = 0.659$ to fit this value. The fast RNG are similar to the ones in 4.d, but here the predictability excess is estimated one order of magnitude smaller. Anyway, the precise numerical value of $q_{set}$ is irrelevant, for it is surely much smaller than $q_m$, so that the locality loophole is clearly closed. The p-value is 0.039, much larger than in 4.d and 4.e because of the smaller size of the statistics, but sufficient to close the memory loophole too.

**5. Discussion and Conclusions.**

The Table shows that, leaving aside some details, the five experiments reach their goals. Now I discuss the details worth mentioning.

The Reader may have perceived my admiration regarding the realization of the RNG, especially in the case 4.e. Yet, the condition $q_{set} < q_m$ is not fulfilled with a margin as wide as could be expected in the experiments using the EI. The cause of this weakness is not in the RNG, but in the low values of $J_m$ attained, which are one to four orders of magnitude below the ideal. The same occurs with $P^+(a)$ excepting in 4.a.

Even though the locality loophole is closed according to the established criterion, the mentioned differences and the sensitivity of the results to numbers that are all close to zero leave a sense of uncertainty. The excellent value of S obtained in 4.c with the same setup indicates that CHSH is more robust than EI and should be preferred, when possible. The reason why EI is chosen in the experiments 4.b-4.e is that the detectors' efficiencies are measured lower than $\eta_{thr}$ for CHSH.

The results of 4.f are more satisfactory, because they not merely violate the inequality (S>2), but get halfway close to the ideal value $2\sqrt{2}$. In order to achieve this remarkable result, the setup combines the advantages of (propagating) photons and the high detection efficiency of (stationary) NV spins. Entanglement swapping is used to teleport the photons' state to the remote NV spins. Nevertheless, entanglement swapping is a pure QM phenomenon with no classical or semi-classical counterpart. I wonder if a true *logical* loophole might be lurking there. For, in order to select the results of the NV measurements to be included into the statistically relevant set, one must assume QM correct, and QM definitely violates the Bell's inequalities. Perhaps, what is believed to be *demonstrated* true by the observations is what is being *assumed* true from the data selection. This is a subtle issue that deserves to be studied in detail elsewhere.

The time-coincidence loophole was closed few years ago [28,29], but not simultaneously with the others. The experiment 4.c claims having closed the detection and the time-coincidence loopholes together. Even though the setup is apparently able to reach this goal, the data analysis provided is, in my view, insufficient. A further analysis [30] uses the method developed in [20-21] and is based on the definition of a "distance" between time series to close the loophole, reaching a p-value smaller than $1.16 \times 10^{-10}$. The achievements of this approach, including the detection of a fake source of entangled states, are remarkable. Yet, I find this approach more complex and indirect than the originally proposed one [2,9] and, in consequence, more vulnerable to new loopholes. I believe that a "traditional" analysis of the time-tagged data in 4.d and 4.e, including the detections *outside* the natural time, is a more reliable way to close the time-coincidence loophole simultaneously with the others. As far as I know this analysis has not been done yet, but it may be done easily in the near future.

An issue that concerns me is the TES jitter, mentioned $\approx 1$ μs in 4.c, and that is increased, in the case of 4.d, because of the use of SQUID amplifiers. This uncertainty may affect the ability of the data produced in these experiments to close the time-coincidence loophole. The reported jitter is larger, at the speed of light, than the distance between the stations, so that even the space-like separation between measurements may be at stake.

Finally, 4.f is truly "event-ready" and hence is not affected by the time-coincidence loophole. Therefore, it closes all the loopholes without further analysis of the data.

Note that the closing thresholds have been calculated for each loophole separately. If the loopholes are combined (for example: $q > ½$ and $\eta < 1$), the thresholds change depending on the details of the hidden variables model. It should also be kept in mind that more loopholes may be found in the future. Faking techniques, which have consequences for the security of QKD schemes, may be considered as a sort of new loophole [31]. It is therefore conceivable that an experiment entirely free of loopholes cannot be done [32]. What real experiments can do, in my opinion, is to shrink the space left to the loophole-based theories down to the point that (say, asymptotically) they become too exotic to be tenable. Also in my personal opinion, the reported experiments have already reached this point, and the defenders of LR should not insist on the little vulnerability left, but to pay attention to some barely explored alternatives. For example, the possibility of non-ergodic dynamics at the hidden variables level [33,34].

Because of its purpose, this report must be critical with the claimed results. Yet, I would like to emphasize my admiration for the extraordinary skills demonstrated by all the groups. They have developed new and formidable abilities of practical interest in the field of quantum information. They have also climbed many steps towards the ideal Bell's proposal. To say the least, they have imposed new and severe restrictions to the set of loophole-based theories.

**Acknowledgements.**

Many thanks to Scott Glancy and Lynden K. Shalm (NIST, Boulder) and to Yanbao Zhang (IQC, University of Waterloo) for their reading of the earlier versions of this paper, their interest, observations and remarks. This contribution received support from the grant PIP11-077 CONICET.

**References.**

TABLE: Summary of some parameters of interest; $J_m$ is the experimentally obtained value of the lhs of the EI, $J_{QM}$ is the QM ideal prediction, $J_{corr}$ is the QM prediction but corrected by the measured efficiencies $\eta_{meas}$, $\eta_{eq}$ is the efficiency value that makes $J_{corr} = J_m$, $q_{QM}$ is the correlation probability between the hidden variable $\mu$ and the analyzers' settings the HV+DZ needs to reproduce $J_{QM}$, $q_m$ is the same but to reproduce $J_m$, $q_{set}$ is the predictability of the RNG as estimated by the authors of the experiments.

|  | Giustina et al. 2013, see 4.b | Christensen et al. 2013, see 4.c | Giustina et al. 2015, see 4.d | Shalm et al. 2015, see 4.e | Hensen et al. 2015, see 4.e. |
|---|---|---|---|---|---|
| $J_m$ | $5.24 \times 10^{-3}$ $\pm 8 \times 10^{-5}$ | $5.4 \times 10^{-5}$ $\pm 7 \times 10^{-6}$ | $7.27 \times 10^{-6}$ | $1.41 \times 10^{-5}$ | $S_m = 2.42 \pm 0.07$ |
| $J_{QM}$ | 0.0701 | 0.0549 | 0.0671 | 0.0645 | $S_{QM} = 2\sqrt{2}$ |
| $J_{corr}$ | $8.53 \times 10^{-3}$ | $5.7 \times 10^{-2}$ | $1.03 \times 10^{-2}$ | $6.26 \times 10^{-3}$ | $S_{corr} = 2.30$ |
| $\eta_{meas}$ | 0.738 and 0.786 | 0.75 | 0.786 and 0.762 | 0.747 and 0.756 | 0.971 and 0.963 |
| $\eta_{eq}$ | 0.745 | 0.710 | 0.719 | 0.715 | not applicable |
| $\eta_{thr}$ | 0.667 | 0.667 | 0.667 (?) | 0.725 | 0.828 |
| $J_{corr}/J_m$ | 1.6 | 106 | 1422 | 444 | 0.95 |
| $q_{QM}$ | 0.785 | 0.787 | 0.783 | 0.776 | 0.769 |
| $q_m$ | ½ + $2.1 \times 10^{-2}$ | ½ + $8 \times 10^{-3}$ | ½ + $7.6 \times 10^{-3}$ | ½ + $1.6 \times 10^{-2}$ | 0.659 |
| $q_{set}$ | 1 | 1 | ½ + $1.2 \times 10^{-4}$ | ½ + $10^{-4}$ | ½ + $10^{-5}$ |
| *p-value* | not computed | $1.16 \times 10^{-10}$ [30] | $3.74 \times 10^{-31}$ | $5.9 \times 10^{-9}$ | 0.039 |
| Coinc.rate (min) | 232 s$^{-1}$ | 1.7 s$^{-1}$ | 9.6 s$^{-1}$ | 0.059 s$^{-1}$ | $\approx 2 \times 10^{-5}$ s$^{-1}$ |
| Coinc.rate (max) | 3970 s$^{-1}$ | 31 s$^{-1}$ | 162 s$^{-1}$ | 3.60 s$^{-1}$ | $\approx 1.3 \times 10^{-4}$ s$^{-1}$ |